\documentclass[11pt, twoside]{article}

\usepackage{ucs}
\usepackage[utf8x]{inputenc}
\usepackage[english]{babel}
\usepackage{amsmath}
\usepackage{amsfonts}
\usepackage{bbm}
\usepackage{bm}
\usepackage{amssymb}
\usepackage{amsthm}
\usepackage{mathtools}
\usepackage{mathrsfs}
\usepackage{graphicx}
\usepackage{algorithmic}
\usepackage{algorithm}
\usepackage{booktabs}
\usepackage{multirow}
\usepackage{fancyhdr}
\usepackage{enumerate}
\usepackage{setspace}
\usepackage{color}

\usepackage[top=2.5cm, bottom=2.5cm, left=1.5cm, right=1.5cm]{geometry}
\usepackage{multicol}
\usepackage{caption}

\usepackage[table]{xcolor}
\usepackage{picture}

\usepackage{natbib}
\usepackage{sectsty}
\allsectionsfont{\large\sffamily}

\floatname{algorithm}{Procedure}

\newcommand{\eps}{\varepsilon}

\newcommand{\prob}{\mathsf{P}}

\newcommand{\ud}{\mbox{d}}

\theoremstyle{plain}

\theoremstyle{definition}

\theoremstyle{remark}

\allowdisplaybreaks[1]

\author{Mat\'{u}\v{s} Maciak, Ostap Okhrin, and Michal Pe\v{s}ta}
\title{Dynamic and granular loss reserving with copulae}

\begin{document}
	
\maketitle

\begin{abstract}
An intensive research sprang up for stochastic methods in insurance during the past years. To meet all future claims rising from policies, it is requisite to quantify the outstanding loss liabilities. Loss reserving methods based on aggregated data from run-off triangles are predominantly used to calculate the claims reserves. Conventional reserving techniques have some disadvantages: loss of information from the policy and the claim's development due to the aggregation, zero or negative cells in the triangle; usually small number of observations in the triangle; only few observations for recent accident years; and sensitivity to the most recent paid claims.

To overcome these dilemmas, granular loss reserving methods for individual claim-by-claim data will be derived. Reserves' estimation is a~crucial part of the risk valuation process, which is now a front burner in economics. Since there is a growing demand for prediction of total reserves for different types of claims or even multiple lines of business, a time-varying copula framework for granular reserving will be established.
\end{abstract}

\noindent JEL classification: C13, C14, C15, C22, C23, C32, C33, C51, C52

\begin{multicols}{2}

\titlepage

\section{Introduction}
Claims reserving is a~classical problem in \emph{non-life insurance}, sometimes also called general insurance (in UK) or property and casual insurance (in USA). A~non-life insurance policy is a~contract between the insurer and the insured. The insurer receives a~deterministic amount of money, known as premium, from the insured in order to obtain a~financial coverage against well-specified randomly occurred events. If such an event (claim) happens, the insurer is obliged to pay in respect of the claim a~claim amount, also known as loss amount.

\emph{Loss reserving} now means, that the insurance company puts sufficient provisions from the premium payments aside, so that it is able to settle all the claims (losses) that are caused by these insurance contracts. The main issue is how to determine or \emph{estimate} these \emph{claims reserves}, which should be held by the insurer so as to be able to meet all future claims arising from policies currently in force and policies written in the past. 

Claims reserving methods based on \emph{aggregated data} from run-off triangles are predominantly used to calculate the claims reserves, see~\cite{EV2002} or~\cite{wutrich_kniha} for an overview. Conventional reserving techniques have some \emph{disadvantages}: loss of information from the policy and the claim's development due to the aggregation; zero or negative cells in the triangle; usually small number of observations in the triangle; only few observations for recent accident years; and sensitivity to the most recent paid claims.

In order to overcome the above mentioned deficiencies or imperfections, \emph{granular loss reserving methods for individual claim-by-claim data} need to be derived. Moreover, estimation of the whole probabilistic distribution of reserves is a~crucial part of the risk valuation process. Since there is a~growing demand for prediction of total reserves for different types of claims or even multiple lines of business, a~\emph{copula framework} for the granular reserving seems to be suitable to handle this multidimensional problem.

\section{State of art and preliminary work}
An example of a~general non-life insurance \emph{claim development} can be illustrated in Figure~\ref{fig:claim-dev}. A~claim occurs at time $t_1$ and is reported with some delay at time $t_2$. Time $t_2-t_1$ is a~\emph{reporting delay}. After some internal process is carried out, an~insurance company pays claim payments at times $t_3$, $t_4$ and $t_5$. It can sometimes happen, that after closing the claim (in time $t_6$) some further facts might emerge and cause re-opening of the claim in time $t_7$ that could lead to revaluation of the claim and further payment in times $t_8$ and $t_9$. Then, the final closure of the claim is in time $t_{10}$. Time between reporting and settling (closing) the claim $t_6-t_2$ or $t_{10}-t_2$ is called a~\emph{settlement delay}.

Because of the reporting and settlement delays, the claims cannot be settled immediately. Therefore, insurance companies need to hold reserves to cover these losses.

We generally distinguish between two types of claims reserves. \emph{RBNS} stands for Reported But Not Settled claims. These are the claims that had occurred and were reported before the present moment, but their settlement would occur in the future. Hence in Figure~\ref{fig:claim-dev}, the present moment is somewhere between $t_2$ and $t_6$ or $t_{10}$. \emph{IBNR} stands for Incurred But Not Reported claims. These are the claims that occurred before the present moment, but will be reported in the future. Therefore, the present moment is somewhere between $t_1$ and $t_2$.

{\centering
\setlength{\unitlength}{.7cm}
\begin{picture}(12,9.5)(-6,-4)
	\thicklines
	\put(-6.0,-0.5){\vector(1,0){12}}
	\thinlines
	\put(-4.5,-0.5){\oval(3.0,0.5)}
	\put(-6.6,-2.2){Insurance period}
	
	\linethickness{0.6mm}
	\put(-4.2,-0.9){\line(0,1){.8}}
	\thinlines
	\put(-4.2,3.2){\vector(0,-1){3}}
	\put(-4.2,-3.0){\line(1,0){1.8}}
	\put(-4.2,-3.1){\line(0,1){0.2}}
	\put(-5.8,3.4){Accident date}
	\put(-3.95,-3.6){IBNR}
	\put(-4.3,-1.5){$t_1$}
	
	\linethickness{0.6mm}
	\put(-2.4,-0.9){\line(0,1){.8}}
	\thinlines
	\put(-2.4,2.2){\vector(0,-1){2}}
	\put(-2.4,-2.5){\line(1,0){7.5}}
	\put(-2.4,-3.1){\line(0,1){0.2}}
	\put(-2.4,-2.6){\line(0,1){0.2}}
	\put(-3.8,2.4){Reporting date}
	\put(0.6,-3.1){RBNS}
	\put(-2.5,-1.5){$t_2$}
	
	\linethickness{0.6mm}
	\put(-1.7,-0.9){\line(0,1){.8}}
	\thinlines
	\put(-1.7,1.2){\vector(0,-1){1}}
	\linethickness{0.6mm}
	\put(-1.4,-0.9){\line(0,1){.8}}
	\thinlines
	\put(-1.4,1.2){\vector(0,-1){1}}
	\linethickness{0.6mm}
	\put(-0.5,-0.9){\line(0,1){.8}}
	\thinlines
	\put(-0.5,1.2){\vector(0,-1){1}}
	\put(-2.2,1.4){Payments}
	\put(-1.85,-1.5){$t_3$}
	\put(-1.4,-1.5){$t_4$}
	\put(-0.6,-1.5){$t_5$}
	
	\linethickness{0.6mm}
	\put(0.4,-0.9){\line(0,1){.8}}
	\thinlines
	\put(0.4,3.2){\vector(0,-1){3}}
	\put(-1.1,3.4){Claim closing}
	\put(0.3,-1.5){$t_6$}
	
	\linethickness{0.6mm}
	\put(2.1,-0.9){\line(0,1){.8}}
	\thinlines
	\put(2.1,2.2){\vector(0,-1){2}}
	\put(1.0,2.4){Reopening}
	\put(2.0,-1.5){$t_7$}
	
	\linethickness{0.6mm}
	\put(2.7,-0.9){\line(0,1){.8}}
	\thinlines
	\put(2.7,1.2){\vector(0,-1){1}}
	\linethickness{0.6mm}
	\put(2.9,-0.9){\line(0,1){.8}}
	\thinlines
	\put(2.9,1.2){\vector(0,-1){1}}
	\put(2.4,1.4){Payments}
	\put(2.5,-1.5){$t_8$}
	\put(2.95,-1.5){$t_9$}
	
	\linethickness{0.6mm}
	\put(5.1,-0.9){\line(0,1){.8}}
	\thinlines
	\put(5.1,3.2){\vector(0,-1){3}}
	\put(5.1,-2.6){\line(0,1){0.2}}
	\put(3.6,3.4){Claim closing}
	\put(5.0,-1.5){$t_{10}$}
\end{picture}

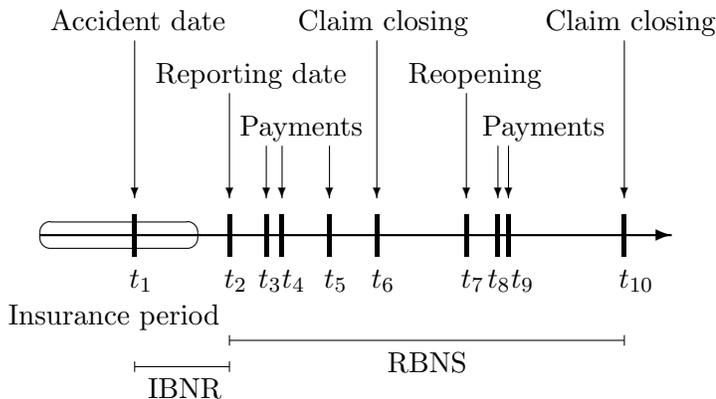
\captionof{figure}{Time development of a~non-life insurance claim.}\label{fig:claim-dev}
}
\vspace{.3cm}

To estimate the distribution of IBNR and RBNS reserves means to predict future cash flows and their uncertainty. On the top of that, this will become compulsory due to the introduction of new supervisory guidelines (Solvency~II).

The loss reserving approaches based on individual/claim-by-claim/micro-level/granular data do not represent a~mainstream in the reserving field. First attempts within the reserving framework of incorporating the claim information for reporting delays were using a~bayesian approach~\citep{Jewell1989,Jewell1990}. Substantial branch of the individual loss reserving methods, that are based on a~position dependent marked Poisson process, involves work of \cite{Arjas}, \cite{Norberg,Norberg1999}, \cite{HaastrupArjas}, \cite{Larsen}, and~\cite{AntonioPlat}. Besides that, \cite{ZhaoZhouWang} and \cite{ZhaoZhou} propose semiparametric techniques from survival analysis. Several case studies of the individual reserving approaches can be found in~\cite{TaylorMcGuireSullivan}.


Evaluation of the reserving risk is one of the front burners in actuarial science. Regardless of its importance, practical reserving techniques often forfeit diagnostics of the theoretical models' assumptions. \cite{PH2012} derived sufficient and necessary conditions for appropriateness of the widely used chain-ladder reserving method only few years ago. These conditions can unambiguously distinguish whether the chain-ladder provides consistent estimates or not. Moreover, the majority of the classical reserving approaches are based on various assumptions of independence, which can sometimes be unrealistic or at least questionable. The application of generalized estimating equations (GEE) for the reserves' prediction was proposed by~\cite{PH2013}, where the claim amounts within the same accident year are dependent. Various correlation structures were introduced within the GEE framework and model selection criteria were suggested. Another relaxation of some restrictive assumptions in the traditional reserving methods---independent claim amounts and large number of parameters depending on the number of observations---was performed by~\cite{PO2014}. Overcoming previously mentioned pitfalls contributes to the increase in precision of the reserves' prediction (e.g., smaller variability, lighter tails) and allows straightforward claims forecasting beyond the last observed development period.

In order to further extend research started by~\cite{PH2013} and~\cite{PO2014}, one needs an appropriate model for multivariate dependence. Despite of the booming research on copulae, currently there are very few specifications of a~flexible and parsimonious model in higher dimensions with convenient estimation. This problem was tackled in~\cite{okh_okh_schmid_2013a}, who analyzed hierarchical Archimedean copulae (HAC) in details and provide a~multistage estimation procedure. 
This research has been further extended by the allowance of the iterative estimation of the highly-parameterized time series models, cf.~\cite{hau_okh_ris_2015}. These basic papers allows for further developments in the time-varying concept, while observing changes of the structures over the time. 
Moreover, a~dynamic character of the copula can be investigated as suggested by~\cite{GOPV2017}.


\subsection{Motivation and data structure}
Nowadays, modern databases and computer facilities provide a~foundation for loss reserving based on individual data. There is no more reason to rely on the reserving techniques using only aggregated data.

We are in close collaboration with the Czech Insurer's Bureau. It allows us to use the database from its Guarantee Fund for car insurance. The data consist of claims developments from the beginning of 2000 and are continuously updated. Each record in the data set contains:
\begin{itemize}
\setlength\itemsep{-.1cm}
\item Claim ID. If one claim is associated with more payments, each payment is on a~separate row.
\item Type of claim. It can be either \emph{bodily injury} or \emph{material damage}.
\item Accident date (occurrence).
\item Reporting date (notification).
\item Date of payment. It is a~date when the payment is credited to the client's bank account.
\item Amount of payment.
\end{itemize}
In return for providing the data, the Czech Insurers' Bureau (\v{C}KP) will certainly benefit from our results. For example, the member insurance companies of the bureau will acquire more precise methodology for loss reserving pre-tested on case studies.

\section{Objectives and aims}\label{sec:aims}
To investigate current individual claims reserving techniques from a~mathematical and statistical point of view belongs to one of the main aims of the paper. The proper understanding of those methods will also help us not only to compare the available methods, but also to propose new alternative methods.

\subsection{Main goals}
One of the main goals of the paper is to derive new \emph{claim-by-claim} methods for loss reserving in non-life insurance, which provide \emph{consistent data-driven} predictions of the future claims (their frequency and severity) together with full probabilistic distribution without some ad-hoc corrections or so-called expert judgement.

{\bf The first part} of the paper is designed to further \emph{develop granular loss reserving methods}. Indeed, inventing new stochastic methods for loss reserving based on claim-by-claim data becomes of the utmost importance in non-life insurance, because there is no loss of information caused by the old-fashioned data aggregation.

{\bf The second part} will cover \emph{modeling the dependencies} among different types of claims \emph{using the copulae}. Applying the already well known as well as the newly derived copula results and concepts to the developed individual reserving methods tackles the risk valuation process.

\subsection{Employed methods}\label{subsec:model}
To outline the \emph{stochastic approach} of granular loss reserving, we propose a~class of models to predict the future loss payments.

\subsubsection*{Phase 1 -- Accident date}
Let us denote the accident date of the $i$th claims ($i=1,2,\ldots$) by $T_i$ and assume without loosing of generality that these accident dates are chronologically ordered such that $T_{i_1}\leq T_{i_2}$ for $i_1<i_2$  and $T_0\equiv 0$. The \emph{date differences} $V_i=T_{i}-T_{i-1}$ between two consecutive accident dates are supposed to be independent and their conditional distribution given the beginning of the time period $T_{i-1}=t$ is $G_t$, i.e.,
\[
\prob\left[V_i\leq x\big|T_{i-1}=t\right]= G_t(x).
\]
Since we deal with approximately $55{,}000$ claims within years 2000--2016, it is natural to assume that~$G_t$ is Poisson, negative binomial, or their zero-modified versions, cf.~Figure~\ref{fig:datediff}. The database of claim developments provides sufficient information such that the parameters of distribution~$G_t$ can be directly estimated using traditional approaches like maximum likelihood and its robustness will be investigated.

\subsubsection*{Phase 2 -- Reporting delay}
The \emph{reporting delay} (waiting time) of the $i$th claim occured at fixed time $t$ is denoted by $W_i(t)$, which is the time difference between the \emph{occurrence epoch} (accident date) and the \emph{observation epoch} (reporting date) given the occurrence time is fixed to $t$. For simplicity of the analysis, at first step processes $\{W_i(t)\}_t$ are supposed to be independent and identically distributed (iid) for all $i$ and the conditional distribution of the reporting delay $W_i$ with the beginning at the random accident date $T_i$ given the beginning of the time period $T_i=t$ is $H_t$, i.e.,
\[
\prob\left[W_i(T_i)\leq w\big|T_i=t\right]=H_t(w).
\]

{\centering
	\includegraphics[width=\columnwidth]{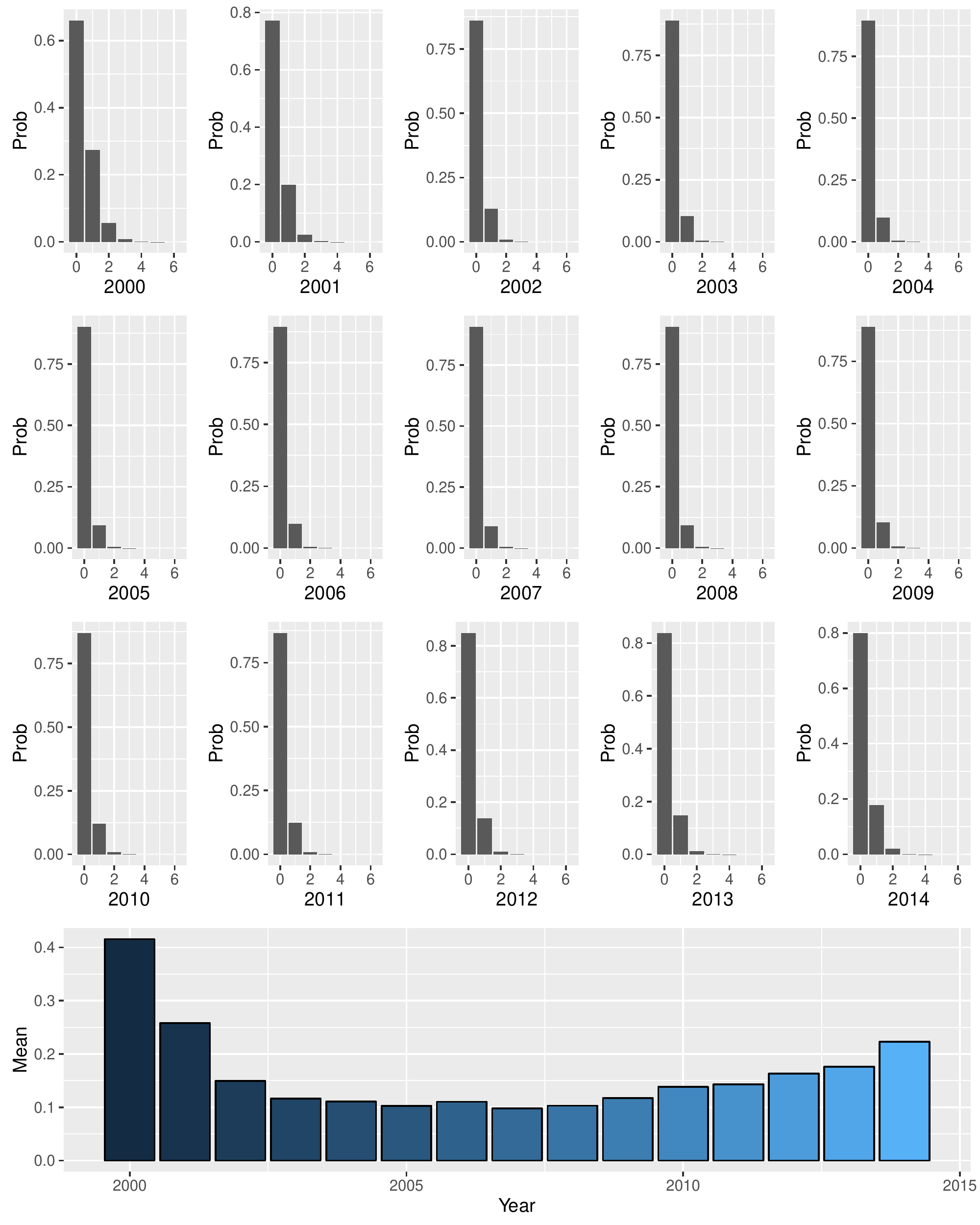}
	\captionof{figure}{Histograms of the date differences for bodily injury by years together with the year-by-year means of the date differences.}\label{fig:datediff}
}
\vspace{.2cm}

For instance, a~Weibull distribution with time-varying parameters can be used for the conditional cumulative distribution function (cdf) $H_t$. The reasoning, why the conditional distribution of reporting delays should depend on time, comes from the exploratory analysis of the claims' database. Figure~\ref{fig:delays} reveals the time effect of accident year and particular day of the year on the reporting delay.

The reporting delays are becoming shorter and shorter, which can be explained by a~possibility of reporting an~accident over internet and even by a~denser net of insurance company's branches. Therefore, we restrict $H_t$ in the way that the conditional expectation of $W_i(T_i)\big|T_i=t$ should be decreasing in $t$. Conditional distribution $H_t$ can be estimated by proposing some parametric form or by non-parametric smoothing. Nevertheless, the number of claims in the database is so high that the empirical cdf nicely serves the purpose of finding a~suitable estimate for $H_t$.

{\centering
	\includegraphics[width=\columnwidth]{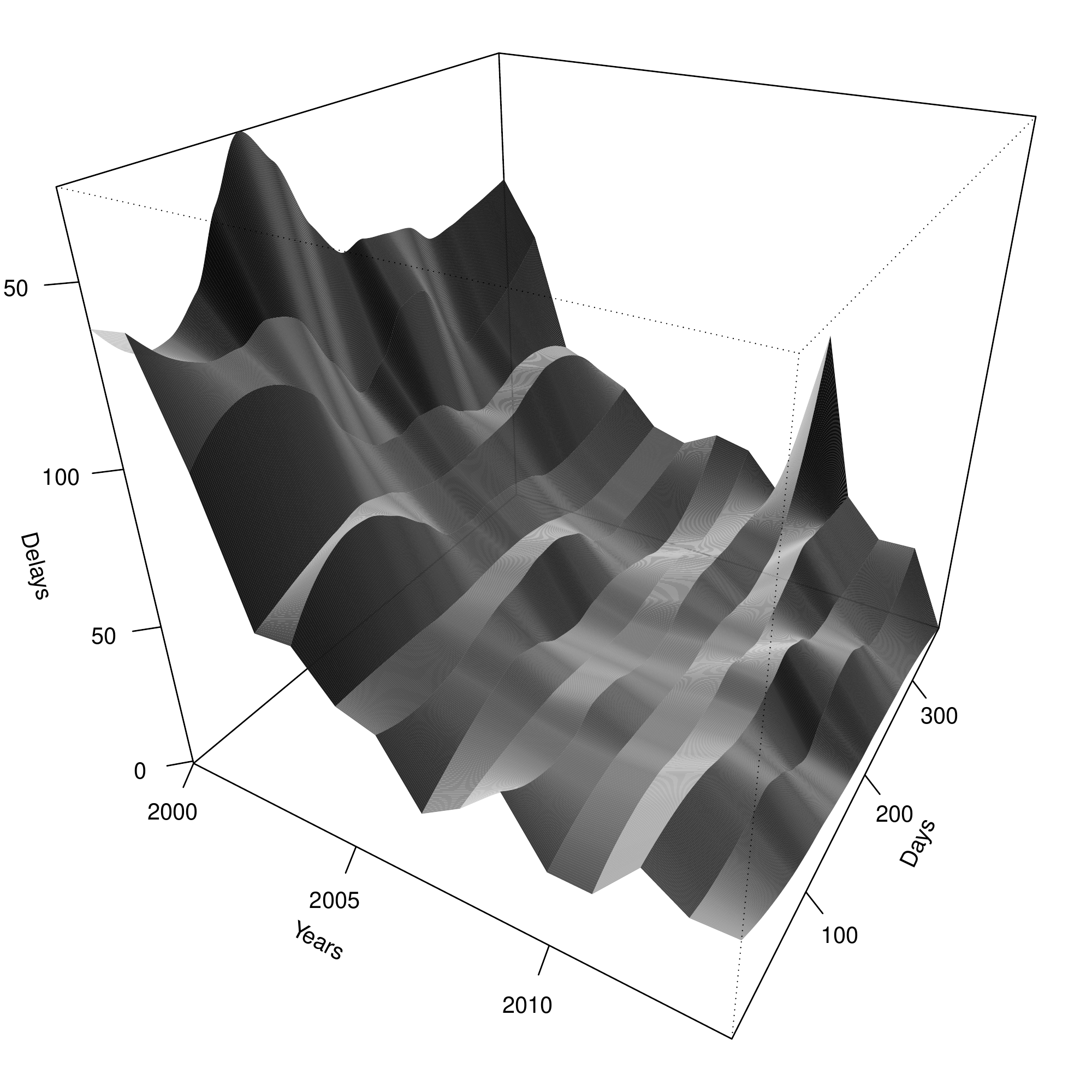}
	\captionof{figure}{Reporting delays (daily medians) for different accident years and accross days within the accident year.}\label{fig:delays}
}
\vspace{.2cm}


\subsubsection*{Phase 3 -- Number of payments}
Suppose that $N_i(\tau)$ represents the number of loss payments corresponding to the $i$th claim during time $\tau$ after the reporting date, i.e., how many payments were carried out within the time window of length $\tau$ after the observation epoch. Time $\tau$ can be though of an~internal claim's time during which the claim is developed after being reported. In general, $\{N_i(\tau)\}_{\tau>0}$ are counting processes for all $i$ and it is believed that they are iid. The hypothesis of independence and identical distribution is planned to be tested at first based on the empirical data. If this is not the case the theoretical model is going to be adjusted appropriately. An~inhomogeneous Poisson point process with intensity function $\lambda(\cdot)$ can be chosen as a~candidate for such counting process. Its intensity can be consequently estimated from the historical (already reported) claims' occurrences assuming some parametric decay (e.g., polynomial or exponential). This will also give some insight into the structure of payments. Furthermore, one can extrapolate the intensity function beyond the longest claim's development using decreasing decays having limit~$0$ in infinity.

Since the behavior of insurance company in order to close a~claim has not changed over time and empirical study from the database does not show the time effect of accident date on the number of payments, we assume that $T_i$ is independent of $N_i(\tau)$. This has to be investigated more deeply from the empirical point of view. If any evidence of the time effect occurs, it will be estimated via methods similar to truncated Fourier series in order to obtain time independent observations. The proposed model will also allow for stress-testing in order to test, what happens if, because of some political decisions, strategy of the company changes and how it influences claim reserves. For a~fixed $t>0$, $W_i(t)$ is a~continuous random variable on the positive half of the real line and, for a~fixed $\tau>0$, $N_i(\tau)$ can be viewed as a~discrete random variable. The dependence between the reporting delay and the number of payments is modeled by time-varying copula $C_{\tau}$, which is firstly assumed to be independent of $t$. Using Sklar's theorem, we have
\begin{multline*}
\prob\left[W_i(t)\leq w,N_i(\tau)\leq n\big|T_i=t\right]\\
=C_{\tau}\left(H_t(w),\sum_{k=0}^n\left[\int_0^{\tau}\lambda(s)\ud s\right]^k\frac{\exp\{-\int_0^{\tau}\lambda(s)\ud s\}}{k!}\right).
\end{multline*}
Recall that $C_{\tau}$ is not unique, because of the discrete margin. This allows for deeper investigation of the probabilistic properties of the model, types of the copula allowed for modeling, estimation methods, etc. On one hand, copula $C_{\tau}$ intentionally depends on~$\tau$, because stronger dependence between the reporting delay and the number of payments is expected for a~shorter time horizon of length $\tau$. On the other hand, we do not assume time effect of occurrence time $t$ on the copula $C_{\tau}$. A~parametric time-varying copula can be used for $C_{\tau}$, i.e., $C_{\tau}(\cdot,\cdot)=C\left(\cdot,\cdot;\vartheta(\tau)\right)$. Different copula models are planned to be considered in the simulation and empirical studies. If necessary, new copula models will be developed from the existing ones in order to match dependency properties observed in the data. Understanding the structure of this probability is crucial in order to model state of the claims reserves at given time point (simulation based). In case of observing systematic behavior of the dependency parameter $\vartheta$, time-varying copula models in the spirit of \cite{Patton2006} will be implemented.

Let us assume that the actual (present) time is $a$ and the aim is to model the number of payments in the future time window $(a,b]$. In order to stochastically model the number of payments within time horizon $(a,b]$, one needs to distinguish two cases according to the claim development (cf.~Figure~\ref{fig:claim-dev}):
\begin{itemize}
\setlength\itemsep{-.1cm}
\item reported but not settled (RBNS) claims,
\item incurred but not reported (IBNR) claims.
\end{itemize}

In the \emph{RBNS} case ($a\geq T_i+W_i(T_i)$), realizations of $T_i$ and $W_i(T_i)$ are observed. Suppose that $T_i=t$ and $W_i(T_i)=w$. Then, the conditional probability of the number of payments for the $i$th already reported claim, given that the number of payments up to time moment $a$ is $k$, is modeled as
\begin{align*}
&\prob\left[N_i(b-t-w)-N_i(a-t-w)=n\big|N_i(a-t-w)=k\right]\\
&=\prob\left[N_i(b-t-w)-N_i(a-t-w)=n\right]\\
&=\left[\int_{a-t-w}^{b-t-w}\lambda(s)\ud s\right]^n\frac{\exp\left\{-\int_{a-t-w}^{b-t-w}\lambda(s)\ud s\right\}}{n!}
\end{align*}
for $n\in\mathbbm{N}_0$, because of independent increments of the inhomogeneous Poisson point process. Information of this probability, which is to be estimated and properties of the estimates investigated, is crucial for predicting the number of payments based on the past behavior.

In the \emph{IBNR} case ($a<T_i+W_i(T_i)\leq b$), $T_i$ and $W_i(T_i)$ are not observable. If $T_i+W_i(T_i)>b$, then no payment could appear in $(a,b]$. Moreover, a~payment can only be proceeded if the claim has already been reported. 

Firstly, let us derive the conditional joint density of $[W_i(T_i),N_i(\tau)]\big|T_i$ in the similar fashion as \cite{KBSC2013}, where $Q_{\tau}$ stands for the cdf of $N_i(\tau)$:
\begin{align}\label{eq:condWN}
&\prob\left[W_i(T_i)=w,N_i(\tau)=n\big|T_i=t\right]\nonumber\\
&=\frac{\partial}{\partial w}\Big\{\prob\left[W_i(t)\leq w,N_i(\tau)\leq n\big|T_i=t\right]\Big.\nonumber\\
&\quad\Big.-\prob\left[W_i(t)\leq w,N_i(\tau)\leq n-1\big|T_i=t\right]\Big\}\nonumber\\
&=\frac{\partial}{\partial w}C_{\tau}\left(H_t(w),Q_{\tau}(n)\right)-\frac{\partial}{\partial w}C_{\tau}\left(H_t(w),Q_{\tau}(n-1)\right).
\end{align}
Besides the practical goal of this paper, we also plan to study theoretical properties of the \emph{mixed copulae} (i.e., copulae having continuous as well as discrete margins) like non-uniqueness or smoothness.

Thus, the conditional distribution of the number of payments for the $i$th not reported claim, given its accident date, its reporting delay, and the fact that it will be reported before the end of time horizon $b$, is
\begin{align}\label{eq:WiTi}
&\prob\Big[N_i(b-T_i-W_i)-N_i(a-T_i-W_i)=n\Big.\nonumber\\
&\qquad\Big.\big|T_i=t, W_i(T_i)=w, a<T_i+W_i(T_i)\leq b\Big]\nonumber\\
&=\prob\Big[N_i(b-t-w)=n\Big.\nonumber\\
&\qquad\Big.\big|T_i=t, W_i(T_i)=w, a<T_i+W_i(T_i)\leq b\Big]\nonumber\\
&=\frac{\prob\left[N_i(b-t-w)=n\big|T_i=t,W_i(t)=w\right]}{\prob\left[a<T_i+W_i(T_i)\leq b\right]}\nonumber\\
&=\frac{\prob\left[W_i(t)=w, N_i(b-t-w)=n\big|T_i=t\right]}{H_t'(w)\prob\left[a<T_i+W_i(T_i)\leq b\right]}\nonumber\\
&=\frac{1}{H_t'(w)\prob\left[a<T_i+W_i\leq b\right]}\nonumber\\
&\qquad\bigg\{\frac{\partial}{\partial w}C_{b-t-w}(H_t(w),Q_{b-t-w}(n))\bigg.\nonumber\\
&\qquad\bigg.-\frac{\partial}{\partial w}C_{b-t-w}(H_t(w),Q_{b-t-w}(n-1))\bigg\}
\end{align}
for $a<t+w\leq b$ and $n\geq 1$, because $N_i(a-t-w)=0$ for the IBNR case. Derivative of the copulae in~\eqref{eq:WiTi} can be calculated 
using
\[
\frac{\partial Q_{\tau}(n)}{\partial \tau}=\frac{\lambda(\tau)}{n!}\left[\int_0^{\tau}\lambda(s)\ud s\right]^n\exp\left\{-\int_0^{\tau}\lambda(s)\ud s\right\}.
\]

Such model is planned to be investigated deeply from a~theoretical perspective: properties of the estimates, asymptotics; as well as from an~empirical one: robustness with respect to the assumed copula model, sensitivity with respect to the homogeneous or inhomogeneous Poisson processes, possible reduction of the number of parameters, etc. As the model contains many parameters, some maximization by parts or iterative likelihood maximization techniques will be adapted to the current model. Members of the team already have an~experience with such approaches, cf.~\cite{hau_okh_ris_2015}. It is possible to extend the proposed approach by assuming dependence of $N_i(\tau)$ on $T_i$, where one can consider the accident date as an~external claim's time. Then, the number of claim payments $N_i(\tau,T_i)$ gives us a~\emph{time-spatial} model, where ``time'' is here the external claim's time and the spatiality is represented by the internal claim's time~$\tau$ that may be though of a~time location from the occurrence of the claim. Hence, a~parametric model for the joint conditional behavior of the reporting delay and the number of payments is
\begin{multline*}
\prob\left[W_i(T_i)\leq x,N_i(\tau,T_i)\leq n\big|T_i=t\right]\\
=C_{\tau,t}\left(H_t(x),\prob\left[N_i(\tau,t)\leq n\big|T_i=t\right];\theta(\tau,t)\right).
\end{multline*}
This model will be also applied in simulation studies with different assumptions and, consequently, validated on the real data.

\subsubsection*{Phase 4 -- Claim amounts}
Let us denote the $j$th payment's amount for the $i$th claim by $X_{i,j}$, where $j=1,\ldots,J_i$. The $X_{i,j}$'s are iid over all $j$'s and $i$'s as well with common cdf $F$. This assumption is based on the empirical analysis of the pairwise relationships between the first, second, and third claim payment's amounts shown in Figure~\ref{fig:payments}, and of course will be discussed more thoroughly. Note that this is the simplest setup for the claim payment's amounts, which can be generalized for instance by assuming underlying regression model, where some covariates (e.g., time or age of the insured person) can be considered. Or one can assume a~kind of autoregression model, where the $j$th payment is modeled via the $(j-1)$st payment
\[
X_{i,j}=\alpha_jX_{i,j-1}+\eps_{i,j},
\]
the first payments $X_{i,1}$'s are supposed to be iid, and $\eps_{i,j}$'s are random disturbances. There are, however, some very payments of a~specific size (not visible in the density plot, nor in the scatterplot), which require in-depth data mining analysis.

{\centering
	\includegraphics[width=\columnwidth]{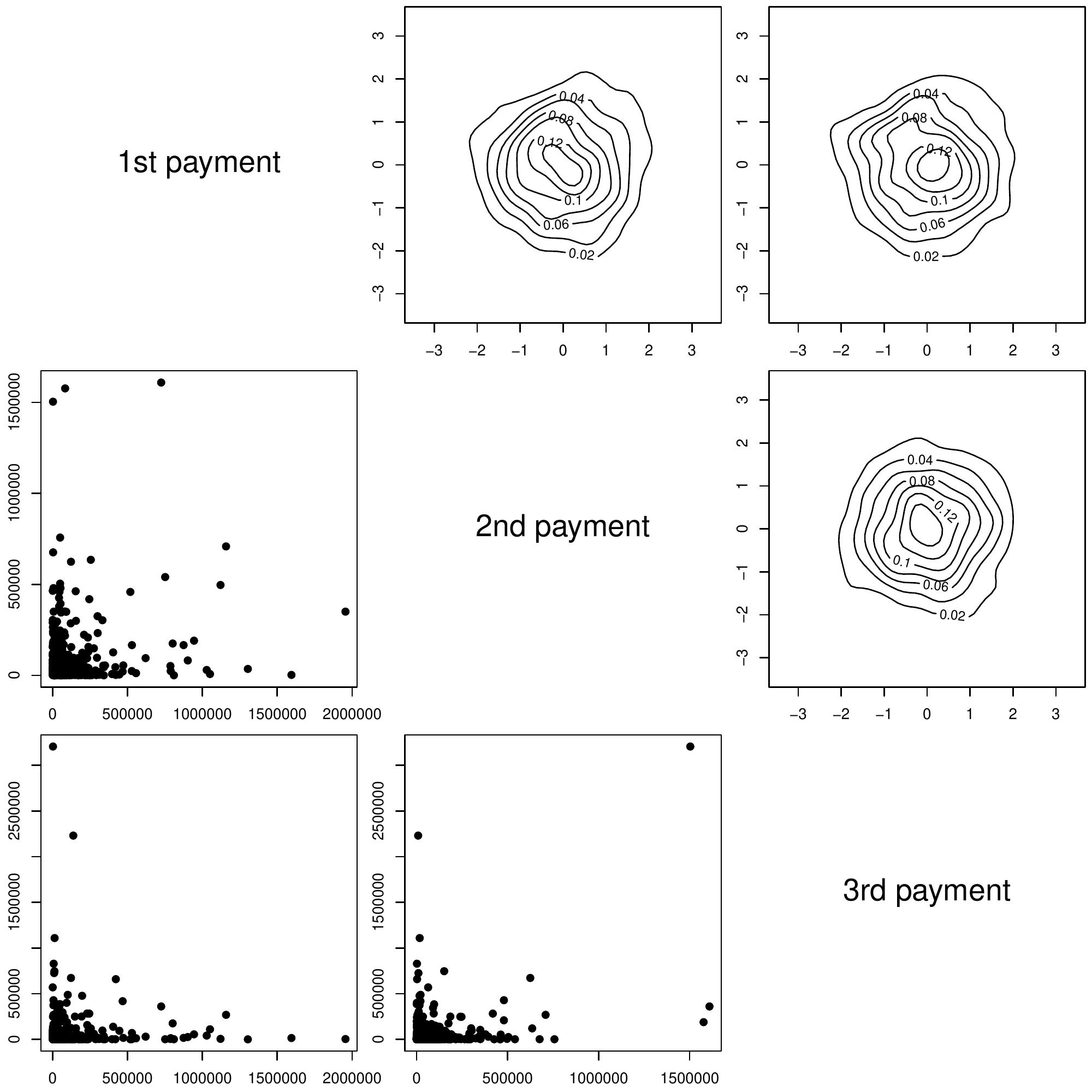}
	\captionof{figure}{Pairwise relationship between claim payment's amounts (bodily injury claims). Subfigures below diagonal show $X_{i,j}$ versus $X_{i,k}$, where $j,k\in\{1,2,3\}$, $j\neq k$. Subfigures above the diagonal display $\Phi^{-1}\{\hat F_j(X_{i,j})\}$ against $\Phi^{-1}\{\hat F_k(X_{i,k})\}$, where $\hat{F}_j$ and $\hat{F}_k$ are the corresponding empirical cdfs and $\Phi$ is the cdf of the standard normal distribution.}
	\label{fig:payments}
}
\vspace{.2cm}

To sum up, identically distributed collections
\[
\left\{T_i,W_i(T_i),N_i(\tau),\{X_{i,j}\}_j\right\}_i
\]
are assumed for the claims. Knowing the stochastic behavior of each component from such collection and their joint relationship, one may predict future claim occurrences and payments, and this is planned to be done in the extensive empirical analysis.


\subsubsection*{Phase 5 -- Several types of claims}
Till this moment, we have restricted our modeling approach only to one type of claim. However, the second type of claim should also be taken into account and can be represented by identically distributed collections
\[
\left\{\tilde{T}_i,\tilde{W}_i(\tilde{T}_i),\tilde{N}_i(\tau),\{\tilde{X}_{i,j}\}_j\right\}_i.
\]
Since there can be two types of claims on one policy, the dependence between two types of claims is modeled by copula~$D$ as
\begin{align*}
&\prob\Big[W_i(T_i)\leq w,N_i(\tau)\leq n,\tilde{W}_i(\tilde{T}_i)\leq \tilde{w},\tilde{N}_i(\tilde{\tau})\leq \tilde{n}\Big.\\
&\qquad\Big.\big|T_i=t,\tilde{T}_i=\tilde{t}\Big]\\
&=D\Big\{C_{\tau}\left(\prob\left[W_i(T_i)\leq w\big|T_i=t\right],\prob\left[N_i(\tau)\leq n\right]\right),\Big.\\
&\qquad\Big.\tilde{C}_{\tilde{\tau}}\left(\prob\left[\tilde{W}_i(\tilde{T}_i)\leq \tilde{w}\big|\tilde{T}_i=\tilde{t}\right],\prob\left[\tilde{N}_i(\tilde{\tau})\leq \tilde{n}\right]\right)\Big\}.
\end{align*}

Moreover, the insurance company has several lines of business (e.g., Motor Insurance or Liability Insurance) and each line of business has its own classes of insurance (corresponding to the types of claims). Hence, the dependence among various lines of business should be modeled as well. This brings us to a~hierarchical structure, where the first level is comprised of the lines of business and the underlying (second) level is formed by the types of claims. Even different groups of the types of claims that are strongly dependent with other types might be grouped on particular levels. Hierarchical Archimedean copulae (HAC) can nicely be used for a~representation of such dependence structure. They are the only copula models that have an~interpretable structure (vine models do not have an interpretation of the structure, as even estimation of the structure does not fall in the class of all vine copula models, but only in some sub-class).

Finally, the above described methods can be generalized by incorporating covariates such as the age of insured person or the region of accident for $X$'s, $W$'s, $T$'s, and $N$'s.


\subsection{Utility for solvency of the insurance company}
The \emph{traditional actuarial view} of reserve risk looks at the uncertainty in the outstanding liabilities over their lifetime. Therefore, the aim is to estimate the \emph{ultimate claims} amount. Although, \emph{Solvency~II} directive view takes a~one year perspective into account, requiring a~distribution of the liabilities after one year.

To summarize all practical issues discussed in previous subsection, assume that we are in time $T=t$ and we want to meet the Solvency~II requirements. If we can model the stochastic behavior of future claims' occurrences, occurrences of incurred but not reported claims, lengths of reporting delay, and the frequency and severity of the loss payments in time, then we possess everything to predict the future cash-flows in time window $(t,t+1\mbox{year}]$ by simulating from the granular loss reserving model. Moreover, one can simulate from the model using the estimated parameters and functionals many times (Monte Carlo based style) in order to obtain simulated distributions of the predictions. Loosely speaking, pushing one button will provide us stochastic predictions for all the future claim payments.

\section{Conclusions}
The paper focuses on three synergic research branches: inventing stochastic methods for loss reserving based on claim-by-claim data, using a dynamic copula framework for modeling dependencies among types of claims, and deriving appropriate statistical inference for these approaches.

\setlength{\bibsep}{0pt}
\setstretch{1}
\bibliography{granular}

\end{multicols}
\end{document}